# Using Excel software to calculate Bayesian factors: taking goodness of fit test (Chi test) as an example

WANG Zhong[1] ✉

(Beijing Doers Education Consulting Co., Ltd, Beijing Doers Institute of Clinical Education)

**Abstract:** Taking the goodness of fit test (Chi test) as an example, this paper attempts to calculate the Bayesian factor BF10 of n-fold Bernoulli test by the Excel software (using JASP software as the evidence). The results showed that in the range of 0.15-0.55 (the rate of samples which are all "true"), the calculated results of Excel were more accurate, and the differences between the two (Excel and JASP) were not statistically significant (P>0.3).

## 1. Introduction

Many literatures have mentioned the limitations of the "null hypothesis significance testing" (NHST). Therefore, many journals now ask to report Bayesian factors (BF) except for p value. There have been many introductions about the limitations of p-value and the advantages of BF, which will not be repeated here. Interested readers can read the articles of Fan Wu (Fan Wu et al., 2018), Chuanpeng Hu (Chuanpeng Hu et al., 2018) and others. However, the calculation of BF is relatively complex (Fan Wu et al., 2018), and it generally depends on software. At present, the more commonly used software is JASP and the R programming language. This leads to a problem: at present, traditional statistical software such as SPSS is more commonly used in biological/medical and sociological (especially psychological) research, but SPSS is not superior in calculating BF, so researchers need to download and learn the use of other software. Although many of these software is free and open source, some of them (such as R) need to learn code rules. This aroused our pure curiosity: Can we use common office software such as Excel to calculate BF?

We will take the chi-square test method (hereinafter referred to as "Chi test") in goodness of fit test as an example to discuss the feasibility of Excel for BF calculation. The Chi test is chosen, on the one hand, because the fitting test is one of the most important statistical ideas in social statistics. Taking structural equation analysis as an example, the basic idea of this analysis is to try to use the model to fit the real data, so as to judge is it appropriate. The key to good fitting is "goodness of fit index" (Jietai Hou et al., 2021, P8), and the most important index is chi-square-square re value. On the other hand, the calculation of BF, such as t-test and ANOVA analysis, has been introduced in many literatures, while the Chi test is not much introduced. More importantly, the Chi test contains recursive calculation, which is just the strength of Excel software.

## 2. Deconstruct the BF factor

Since the Excel software generally does not have the function of calculating BF directly, we need to deconstruct BF first to understand what core indicators need to be calculated by the software.

According to Fan Wu's article (Fan Wu et al., 2018), BF is generally represented by the symbol $BF_{10}$, and there are:

$$\frac{P(H_1|data)}{P(H_0|data)} = \frac{P(data|H_1)}{P(data|H_0)} \times \frac{P(H_1)}{P(H_0)} \quad ①$$

Where $\frac{P(data|H_1)}{P(data|H_0)}$ is the required BF.

$\frac{P(H_1)}{P(H_0)}$ is the priori probability ratio, which can be set according to existing experience. However, more often, due to the prior distribution are unknown as usual, it is often set to 1 to show that "there is no bias to either side of the alternative hypothesis or the null hypothesis" (Fan Wu et al., 2018).

So, if $\frac{P(H_1)}{P(H_0)} = 1$ ,then the formula ① became:

$$\frac{P(H_1|data)}{P(H_0|data)} = \frac{P(data|H_1)}{P(data|H_0)} = BF_{10} \quad ②$$

In other words, we just need to solve $P(H_0|data)$ and $P(H_1|data)$ respectively.

On the other hand, from the definition of null hypothesis and alternative hypothesis, there is $H_0 \cap H_1 = \phi$, and $H_0 \cap H_1 = \phi$ (all of the sample space). From the nature of conditional probability,

$$P(H_0|x_1, \ldots, x_n) + P(H_1|x_1, \ldots, x_n) = P(H_0 \cup H_1|x_1, \ldots, x_n) = 1$$

(Yici Zhang et al, 2000, P24)。

$$\therefore \qquad BF_{10} = \frac{1-P(H_0|x_1, \ldots, x_n)}{P(H_0|x_1, \ldots, x_n)} \qquad ③$$

That is, we only need to solve the conditional probability $P(H_0|x_1, \ldots, x_n)$ , that will do.

**3. Calculate $P(H_0|x_1, \ldots, x_n)$**

Generally speaking, the prior distribution of the event needs to be known to calculate the conditional probability using Bayesian method. However, the prior distribution of different psychological experiments is likely to be very different, which brings difficulties to the calculation.

Considering that most of the general psychological experiments are n-fold Bernoulli tests, for example, the participants are surveyed with a double-blind questionnaire, then the questionnaire of each participant can be regarded as an independent test. For this reason, Yici Zhang et al (2000) an idea was proposed: if the probability of occurrence of a certain n-fold Bernoulli event A is estimated, the prior distribution of A can be regarded as a uniform distribution on (0,1), that is

$$h(p) = \begin{cases} 1 & p \in (0,1) \\ 0 & Others \end{cases}$$

Where h (p) is the distribution density function of the probability p of occurrence of A, and Yici Zhang et al. gave the calculation method of the probability of occurrence of A in this case:

$$\hat{p} = E[p|x_1, \ldots, x_n]$$
$$= \frac{\sum_{i=1}^{n} x_i + 1}{n+2} \qquad ④$$

Where $x_i$ is the ith observation value, with two possible values of 0 or 1 (Yici Zhang et al., 2000, P151-152). It can be seen that the conditional probability is actually an expectation.

However, it should be noted that the probability p is a point estimation, while the formula ③ requires a conditional probability containing the assumption $\{H_i\}$ (i=0,1). Therefore, we must transform the question into a point estimation before we can solve it with the above method.

For this reason, we set event B: the frequency of the first n tests A is 0.5. Obviously,

$$B_n = \begin{cases} 1 & \text{the frequency of the first n tests A is 0.5} \\ 0 & \text{the frequency of the first n tests A is NOT 0.5} \end{cases}$$

So, if Bn occurs, there are:

$$\frac{1}{n}\sum_{i=1}^{n} x_i = 0.5$$

∴ According to the Cauchy's Convergence Test, if p is the probability of A occurred (and p exists), when $n \to \infty$, ∀ ε> 0:

$$\left|\frac{1}{n}\sum_{i=1}^{n} x_i - p\right| < \varepsilon$$

From the law of Large Numbers of Bernoulli,

$$\lim_{n\to\infty} P\{|p - 0.5| > \varepsilon\} = 0$$

（Yici Zhang, 2000, P124）.

So, we proved that when $n \to \infty$, Bn converges to $H_0$. Then, we will transform the question of whether event A occurs into the question of whether event B occurs.

Now make $r_i = |\frac{1}{i}\sum_{j=1}^{i} x_j - 0.5|$, obviously that $r_i \in [0, 0.5]$. It represents the distance between the frequency of A and 0.5 in the first i test.

Another, set $y_i$ is the ith observation of event B, and there are:

$$y_i = \begin{cases} 1 & r_i = 0 \\ 0 & r_i \neq 0 \end{cases} \quad ⑤$$

So, from formula ④, there is

$$\widehat{p_B} = E[Bn|y_1, \ldots, y_n] = \frac{\sum_{i=1}^{n} y_i + 1}{n + 2}$$

Where $\widehat{p_B}$ is the estimation of the overall probability of occurrence of event B. And because when $n \to \infty$, Bn converges to $H_0$, so $\widehat{p_B} \to P(H_0|x_1, \ldots, x_n)$ .

And because of $y_1, \ldots, y_n$ by $x_1, \ldots, x_n$ is the only decision, so when n is sufficiently large, there is

$$P(H_0|x_1, \ldots, x_n) = \frac{\sum_{i=1}^{n} y_i + 1}{n+2} \quad ⑥$$

Since $y_i$ is a piecewise function, it is too complex to express expressions ③ and ⑥ as a unified analytic expression. This article will not expand here. For all that, because it contains recursive relations, expressions ③ and ⑥ can be easily expressed in Excel, as shown below.

**4. Find the unbiased interval of $r_i$**

Although theoretically we can calculate $BF_{10}$ from formulas ③ and ⑥, in actual operation, because the probability of $r_i$ exactly being 0 is very small, if $y_i$ is 1 only when $r_i$ must strictly be equal to 0, it is likely to cause Excel software to be too strict, and all $y_i$ will be judged as 0, which will lead to the failure of formula ⑥. To prevent this, we need to set an unbiased interval [0, k] for $r_i$, that is, when $r_i \in [0, k]$, it is still regarded as $r_i=0$. Therefore, the value of k must not affect the overall conclusion as much as possible.

Therefore, based on the idea of goodness of fit test, we set:

$G_0$: When other conditions are the same, the Chi test results when $r_i=0$ and $r_i=k$ have significant differences.

Then, the alternative assumptions should be:

$G_1$: When other conditions are the same, there is no significant difference between the Chi test results when $r_i=0$ and $r_i=k$

Because the degree of freedom of the n-fold Bernoulli test generally is *df=2*, it can be seen from the Chi-square table that the following only needs to be proved: $P\{\chi^2 > 3.84|G_0 \text{ established}\} < 0.05$.

So, it can be seen from Pearson's theorem that:

$$\chi^2 = \sum_{i=1}^{df} \frac{(\gamma_i - np_i)^2}{np_i} \quad ⑦$$

among $\gamma_i$ is the actual frequency of the first i occurrence before the event, $np_i$ is the theoretical frequency (Yici Zhang et al., 2000, P188). For this question, *df=2*, $np_i=0.5n$, $\gamma_i=n(0.5\pm k)$. Bring the above conditions into formula ⑦, then:

$$\sum_{i=1}^{2} \frac{(n(0.5 \pm k) - 0.5n)^2}{0.5n} = 3.84$$

Since it contains n, that is, the value of k is not unique, let's make n=200 and try to solve the BF.

Substitute n=200 into the above formula and get $k \approx \pm 0.069$. That is, the unbiased interval of ri should be [0, 0.07) when n=200.

**5. Implementation on Excel software**

Therefore, we write the following program in Excel software:

**Table 1: Implementation of the above algorithm in Excel**

| | A | B | C | D | E |
|---|---|---|---|---|---|
| 1 | Num | $x_i$ | $r_i$ | $y_i$ | $BF_0$ |
| 2 | 1 | 1 | —— | —— | —— |
| 3 | 2 | 1 | ① | ② | —— |
| | | | …… | | |

5.1 Cell meaning:

5.1.1 Column A: the sequence number, which is used to renumber and provide the denominator when calculating $BF_0$.

5.1.2 Line 1: $x_i$, $r_i$, $y_i$, $BF_0$ the meaning sees "3. Calculate $P(H_0|x_1, \dots, x_n)$" for more details.

5.1.3 ①: "=ABS(AVERAGE(B$2:B3)-0.5)" for the meaning of this sentence, please refer to the above the definition of $r_i$, where "ABS" is the absolute value function and "AVERAGE" is the arithmetic average function. Note that the first indicator in the

AVERAGE function must be "B $2" (must contain "$"), otherwise Excel defaults to the average of two adjacent cells, not the average of the first i items.

5.1.4 ② this sentence means that if the value of cell C3 is less than 0.07, it will be assigned a value of 1, otherwise it will be assigned a value of 0. The number 0.07 is the upper bound of the value of k (not necessarily the supremum). Where "IF" is the judgment function of Excel.

5.2 Procedure test

When testing the program, we will find a problem: for the same set of sample values, if we disordered the order of the sample, then the calculated BF value is may not unique. The reason is that our definition of event B contains an implicit meaning of recursion (the value of B is determined of the frequency of A in the first i samples), and the recursive sequence itself is highly sensitive to the order, which causes the difference of BF. For this reason, we arranged all the sample sequence as {1, ..., 1, 0, ..., 0}. The reason for this order is that if all the samples which equal to 1 are at the end (namely {0, …, 0, 1, …, 1}), it is likely to cause a large number of cases where the BF value exceeds 100, resulting in algorithm failure.

## 6. Verification and correction

We took JASP software as the evidence software, and calculated BF, where the total number of samples n=200, when $\sum_{i=1}^{200} x_i = 10, 20, 30, \ldots, 110$. The higher (>110) BF values was not calculated because when it was greater than 110, the BF values given by Excel software has all exceeded and equal to 201. The results of Excel and JASP software are shown in the following table (since more often, we prefer to verify the alternative hypothesis, so here JASP we selected ">Test value" and "$BF_{10}$", the same below):

**Table 2: the results of Excel and JASP**

| The sum of $x_i$=1 | The frequency of occur of A | Excel | JASP |
|---|---|---|---|
| 10 | 0.05 | 27.85714 | 0.005 |
| 20 | 0.1 | 15.83333 | 0.006 |
| 30 | 0.15 | 10.22222 | 0.007 |
| 40 | 0.2 | 7.416667 | 0.008 |
| 50 | 0.25 | 5.733333 | 0.01 |
| 60 | 0.3 | 4.771429 | 0.012 |
| 70 | 0.35 | 3.926829 | 0.016 |
| 80 | 0.4 | 3.297872 | 0.022 |
| 90 | 0.45 | 3.590909 | 0.037 |
| 100 | 0.5 | 6.769231 | 0.084 |
| 102 | 0.51 | 7.782609 | 0.106 |
| 105 | 0.525 | 10.88235 | 0.16 |
| 107 | 0.535 | 13.42857 | 0.222 |
| 110 | 0.55 | 21.44444 | 0.398 |

It is obvious that the results of the softwares are not the same, and the difference is large. The reason for this is probably related to the algorithm we choose (the idea of Yici Zhang et al. that the default prior distribution is uniform, but there are probably more suitable prior distributions). To this end, we need to seek the possibility of converting Excel results into JASP, that is, to correct Excel results.

First, let's look at the regression function of the overall results of the two. As shown in Figure 1-2, the scatter plots of Excel and JASP results are shown respectively, and the dotted line are the regression functions. It can be seen from the regression function that the difference between the two are still large. The most obvious difference was that the Excel results have obvious upward in the range of 0-0.15 (corresponding to $\sum_{i=1}^{200} x_i = 10, 20$). If we draw the scatter diagram of $r_i$, it is not difficult to find the key of the problem (Fig. 3-4): for $\sum_{i=1}^{200} x_i = 10, 20$, because we have previously placed all the samples that A happened (that is, $x_i$=1) at the beginning, this results in all the values of $r_{10}$ and $r_{20}$ being 0.5, and then the values of $r_{20}$ and $r_{40}$ will quickly decline to 0 (because $r_i$ is the arithmetic mean of the first i). Therefore, the essence of this deviation is caused by the order of samples mentioned above ({1, ..., 1, 0, ..., 0}). When $\sum_{i=1}^{200} x_i > 20$, because the number of $x_i$ equal to 0 and 1 are gradually close, the bias effect also gradually disappears. Therefore, we rejected the values in $\sum_{i=1}^{200} x_i \in [0, 30)$ (that is $r_i$>0.15), let's look at the overall regression of the two of this time. As shown in Figure 5-6, it can be seen that the shape of the two regression functions has been closer this time, and the $R^2$ of the two regression functions are greater than 0.7, which is acceptable, indicating that the overall trend of the two functions has been very close.

**Figure 1 the regression result of Excel**

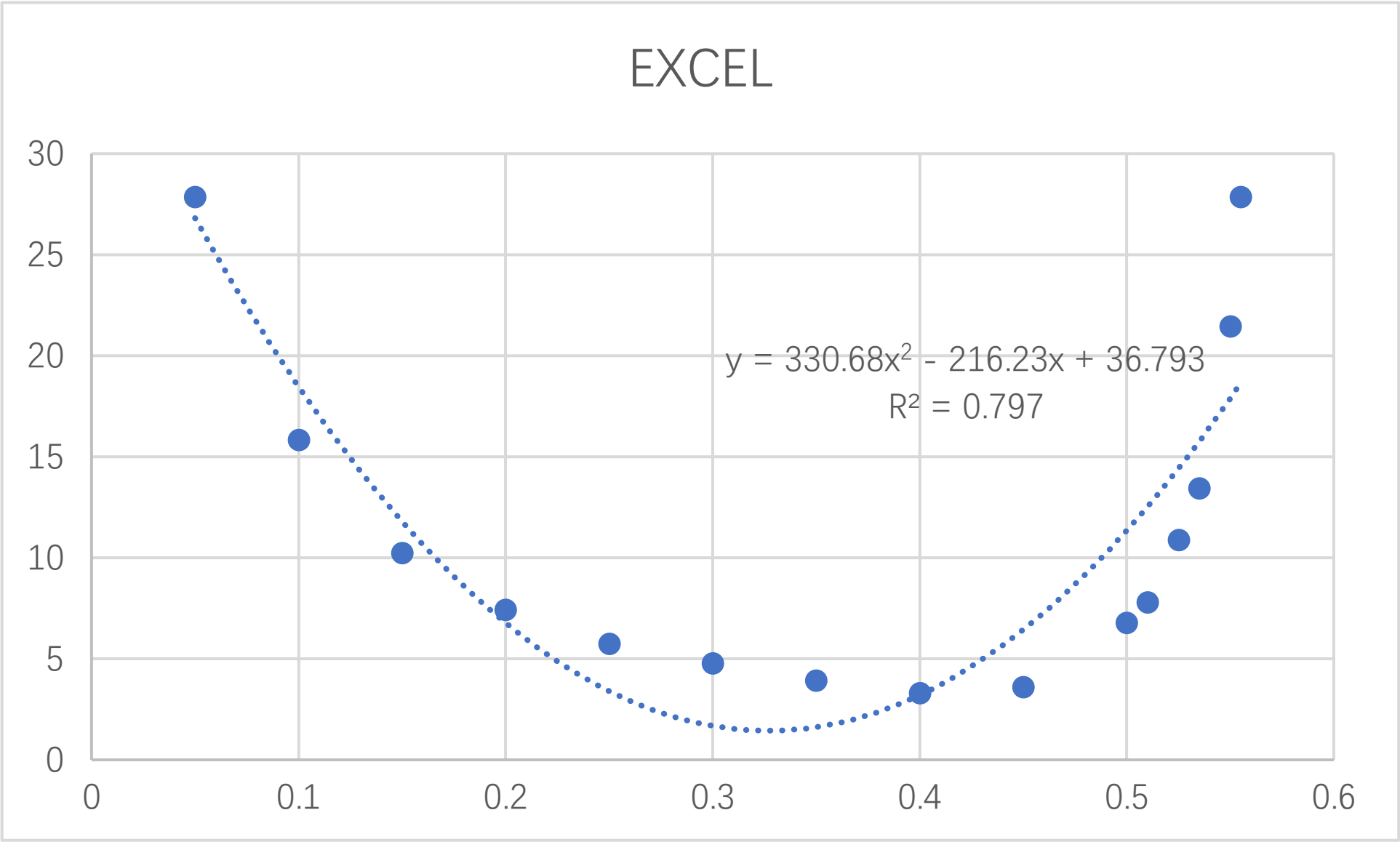

**Figure 2 the regression result of JASP**

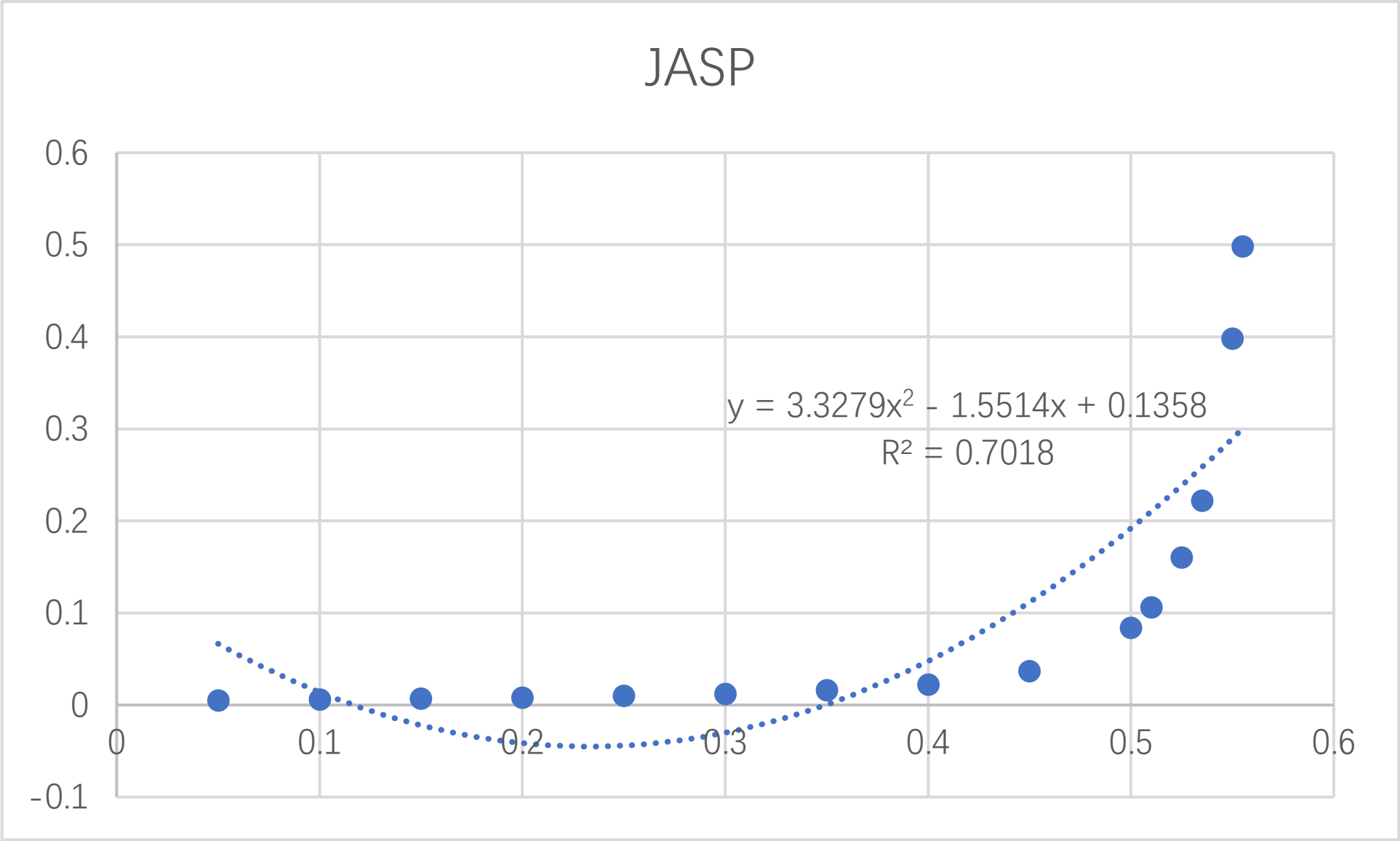

**Figure 3 the value $r_i$ of when $\Sigma x_i$=10 (n=200)**

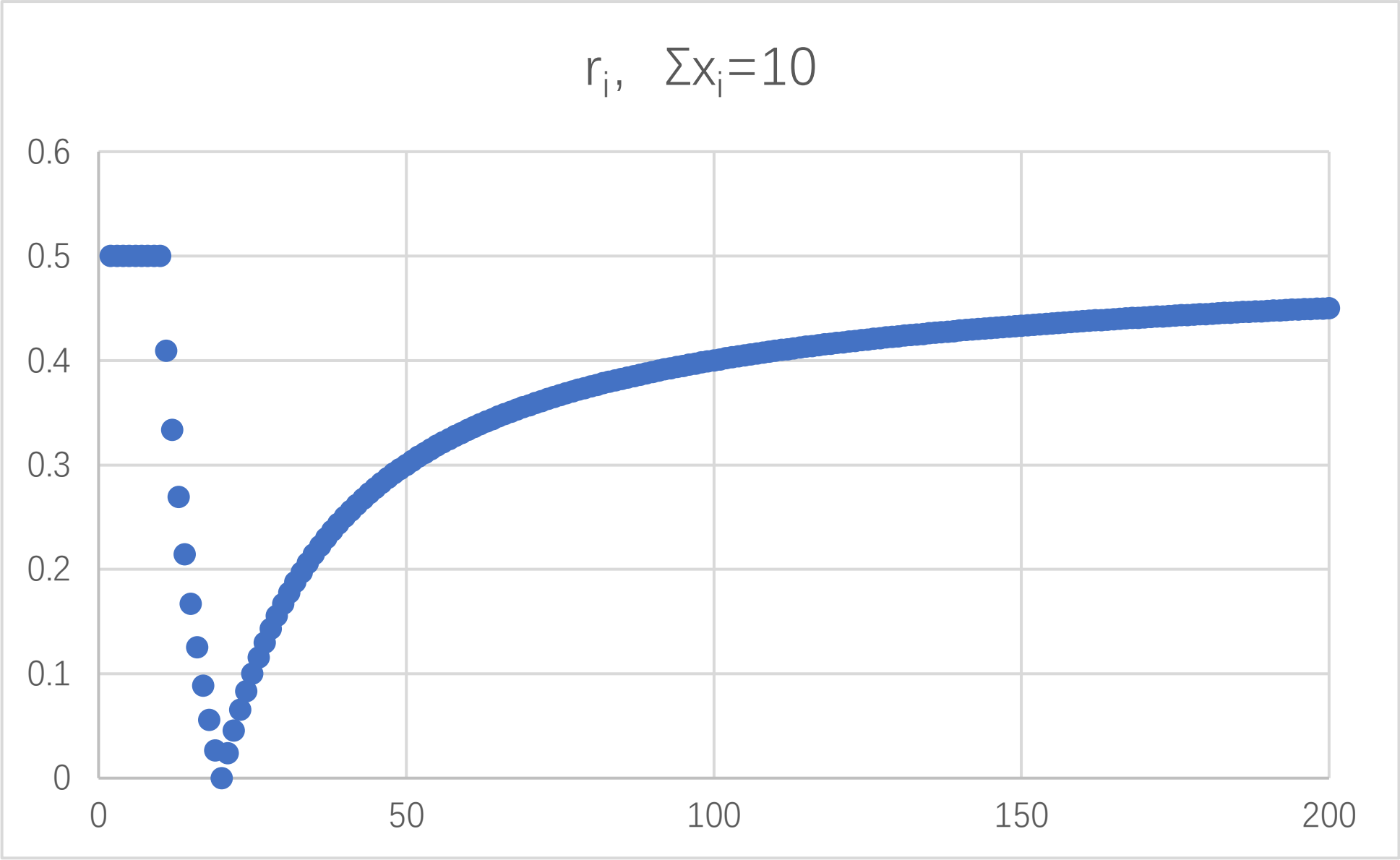

**Figure 4 the value $r_i$ of when $\Sigma x_i$=20 (n=200)**

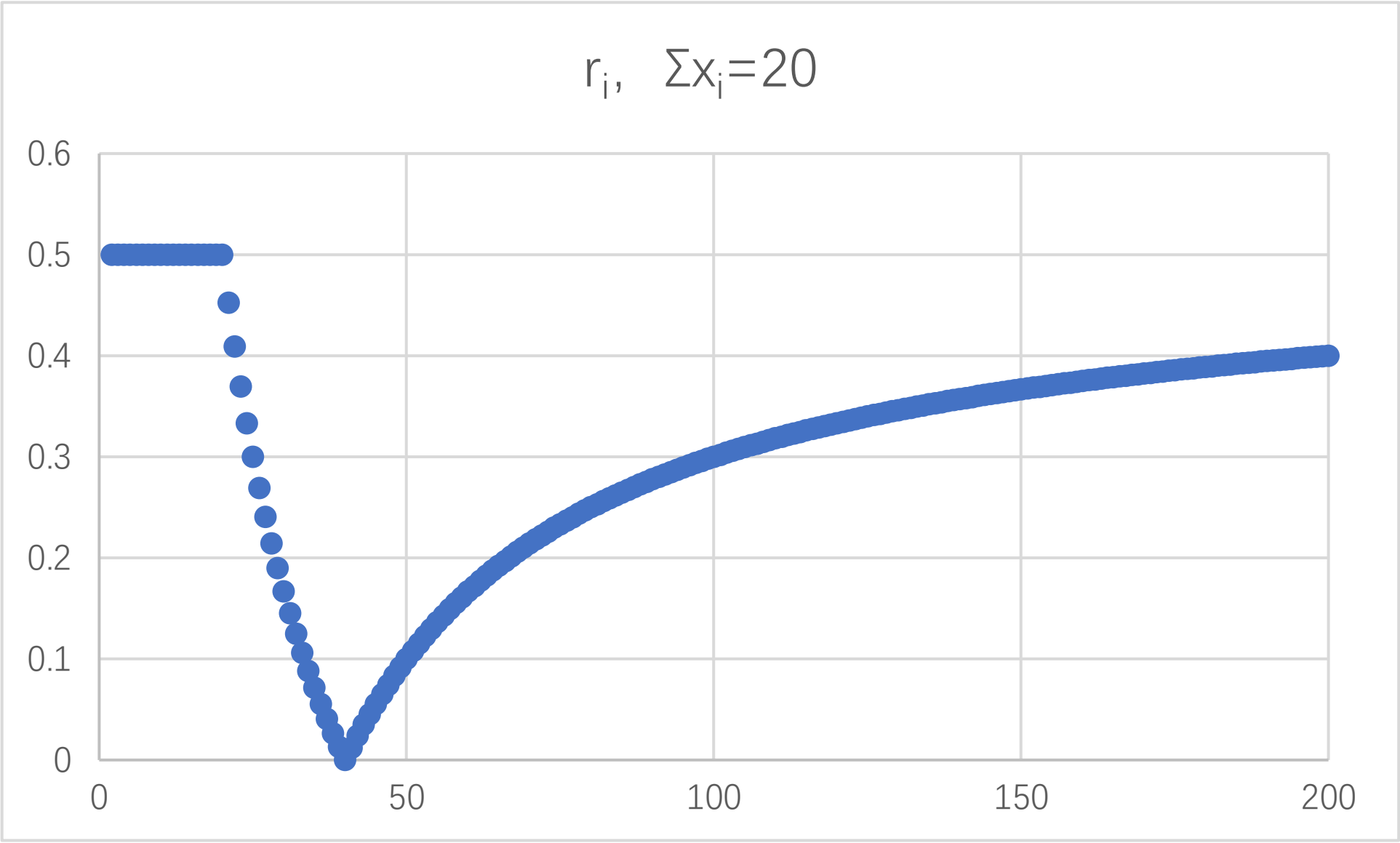

**Figure 5 the regression result of Excel when rejected the value in [0, 30)**

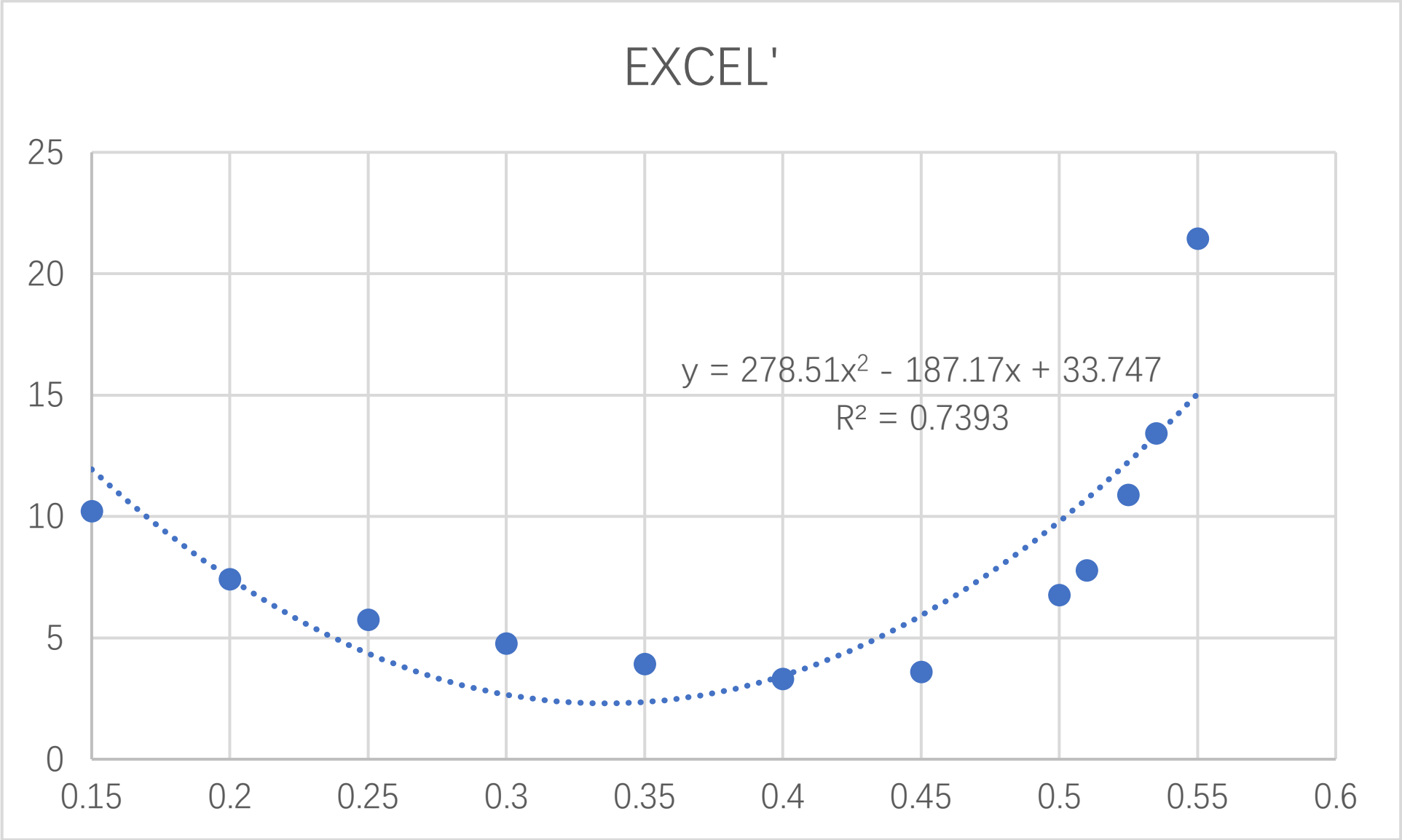

**Figure 6 the regression result of Excel when rejected the value in [0, 30)**

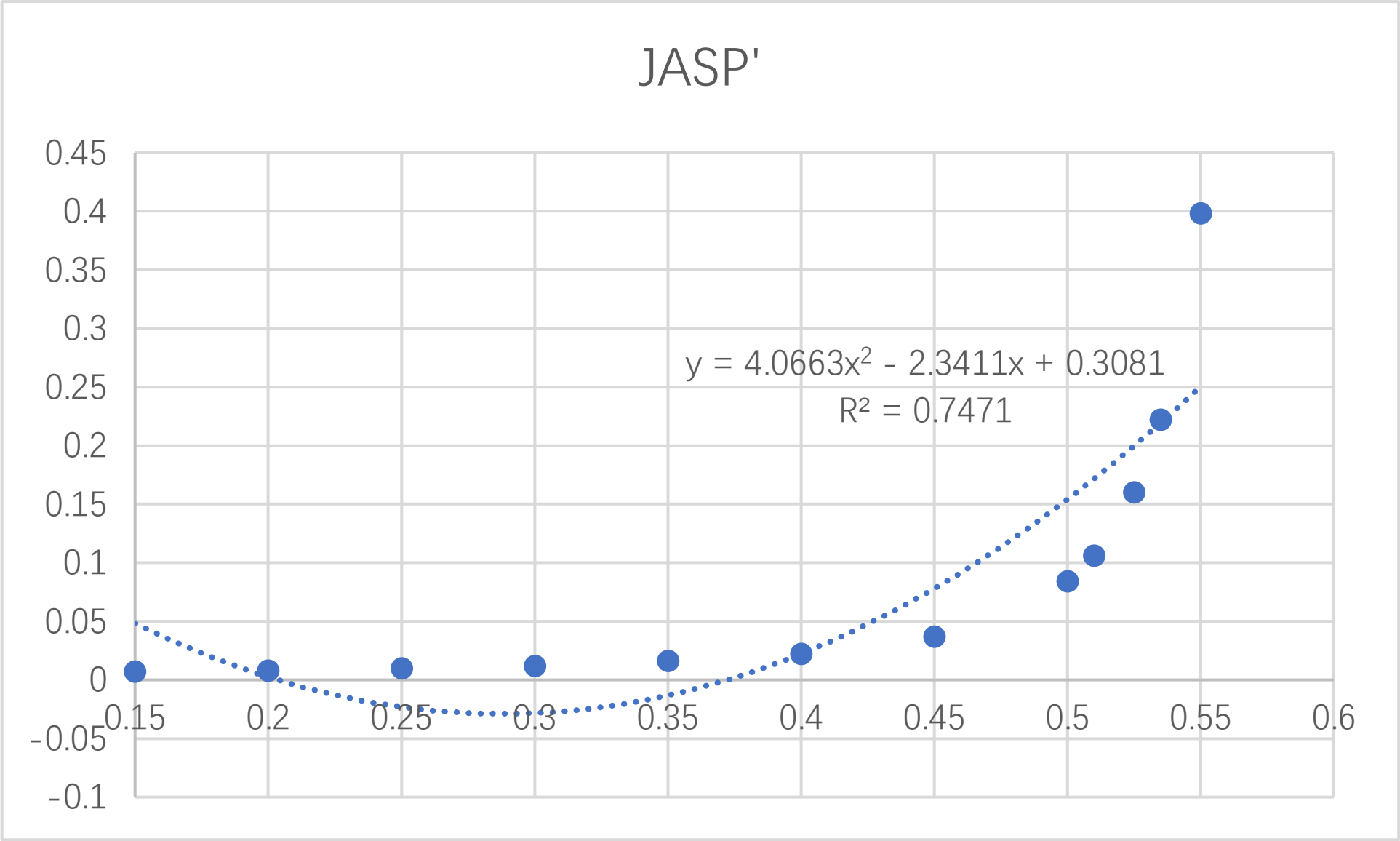

But the problem now is that the $R^2$ of the two is still far less than 0.9, that is, the error is still large. This means that, if the results of Excel are corrected according to these two regression functions at this time, it is likely that there is still a large gap between the results of JASP and Excel. Therefore, we cut the define domains of two regression functions into two parts respectively, 0.15-0.45 and 0.45-0.55, and recalculate their regression functions respectively, which may greatly reduce the error. Figure 7-10 shows the scatter diagram and regression function of Excel and JASP in 0.15-0.45 and 0.45-0.55 respectively. In order to further reduce the error, we used the cubic function to carry out regression. It can be seen that $R^2$ were greater than 0.99, and the fitting were very good.

**Figure 7 the regression result of Excel in 0.15-0.45**

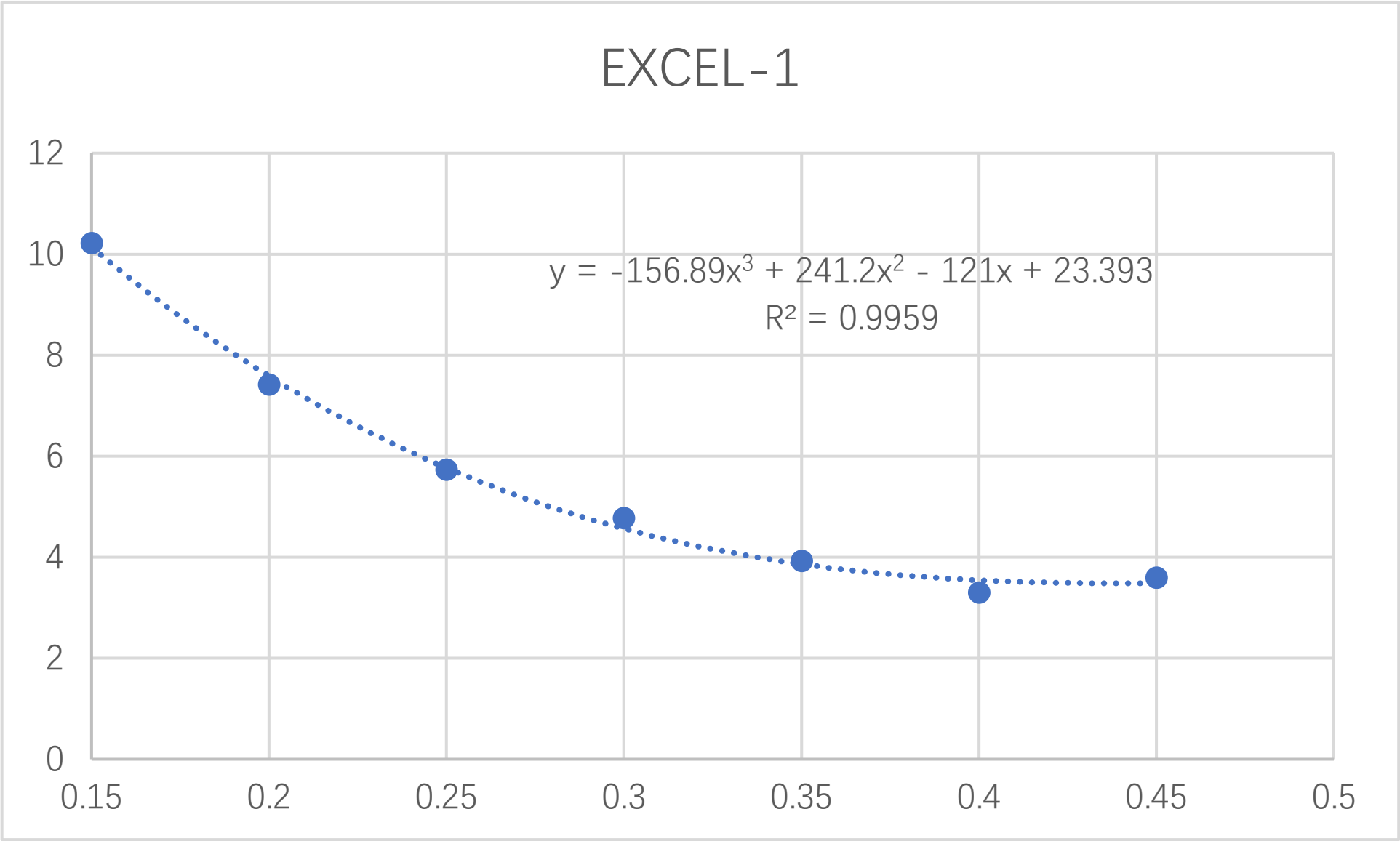

**Figure 8 the regression result of JASP in 0.15-0.45**

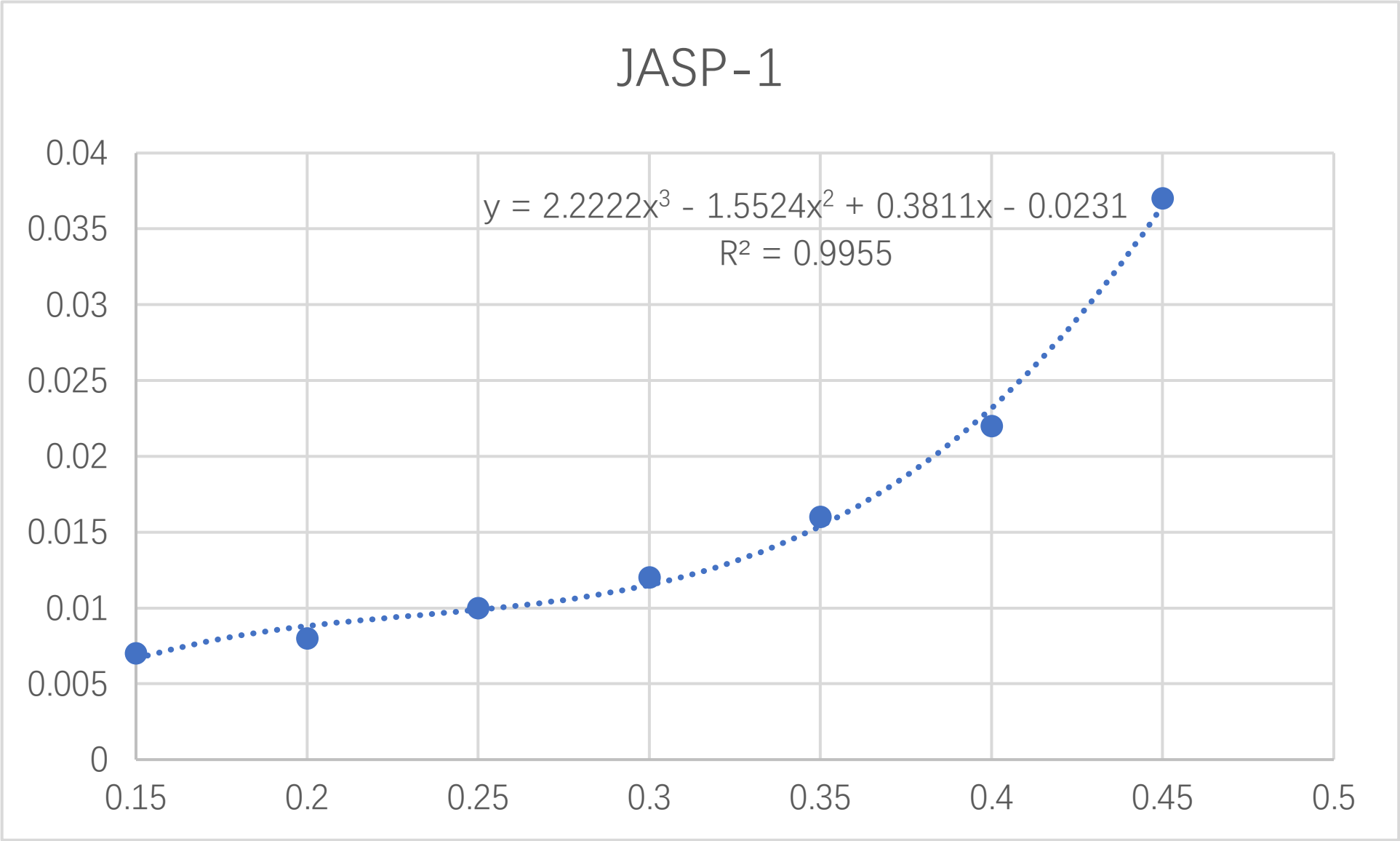

**Figure 9 the regression result of Excel in 0.45-0.55**

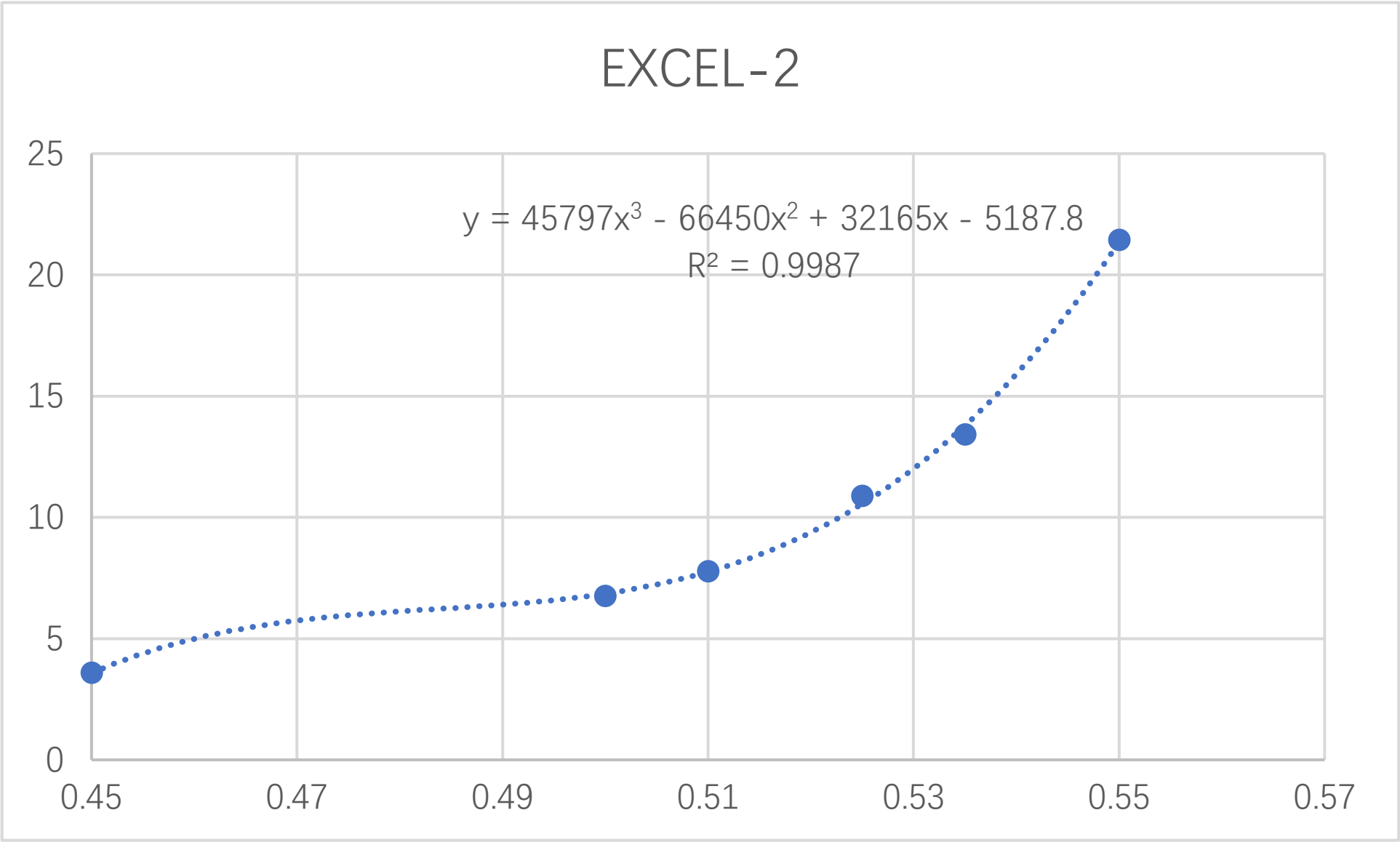

**Figure 10 the regression result of JASP in 0.45-0.55**

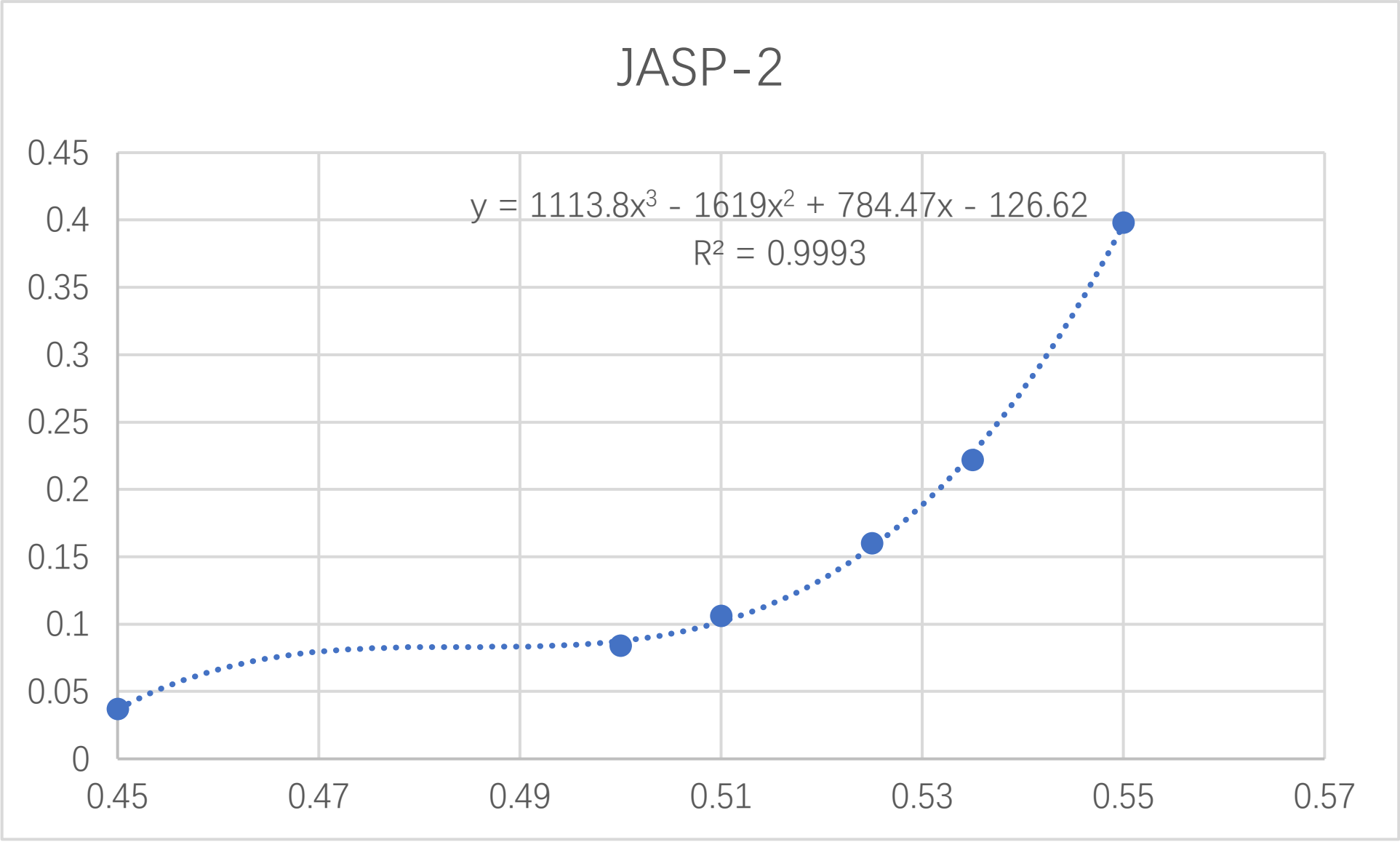

It is also known that if we set two functions:

$$y = a_1x^3 + a_2x^2 + a_3x + a_4$$

$$y = b_1x^3 + b_2x^2 + b_3x + b_4$$

then the above formula can be converted into:

$$b_1x^3 + b_2x^2 + b_3x + b_4 = \frac{b_1}{a_1}y + \frac{a_1b_2 - a_2b_1}{a_1}x^2 + \frac{a_1b_3 - a_3b_1}{a_1}x + \frac{a_1b_4 - a_4b_1}{a_1} \quad ⑧$$

Where $a_1 \neq 0$. That is, the form of upper formula can be transformed as the lower formula through algebraic operation.

∴ Calculate the corrected Excel value from formula ⑧, as shown in the following table (Table 3):

**Table 3 the results of Excel corrected**

| The frequency of occur of A | Excel | Excel corrected | JASP |
|---|---|---|---|
| 0.15 | 10.22222 | -0.00946 | 0.007 |
| 0.2 | 7.416667 | 0.00831 | 0.008 |
| 0.25 | 5.733333 | 0.008857 | 0.01 |
| 0.3 | 4.771429 | 0.02209 | 0.012 |
| 0.35 | 3.926829 | 0.031817 | 0.016 |
| 0.4 | 3.297872 | 0.047038 | 0.022 |
| 0.45 | 3.590909 | 0.068992 | 0.037 |
| 0.5 | 6.769231 | 0.133632 | 0.084 |
| 0.51 | 7.782609 | 0.165269 | 0.106 |
| 0.525 | 10.88235 | 0.289614 | 0.16 |
| 0.535 | 13.42857 | 0.41727 | 0.222 |
| 0.55 | 21.44444 | 1.494681 | 0.398 |

It can be seen that "Excel corrected" and "JASP" were very close. But is that "close" statistically significant? For further confirmation, we used the independent sample t-test method to calculate the significance of the difference between the two groups of data (calculation software: SPSS24). As shown in Table 4, the p value is greater than 0.3, which means that there is no significant difference between the two.

**Table 4 the significant between Excel corrected and JASP**

| | Levene test | | Independent t-test | | | | | | |
|---|---|---|---|---|---|---|---|---|---|
| | F | **Significance** | t | df | **Significance (two tailed)** | Average interpolation | SE interplation | 95% CI of interpolated Lower limit | 95% CI of interpolated Upper limit |
| EV | 3.020 | 0.096 | 1.053 | 22 | **0.304** | 0.13300950 | 0.12627344 | -0.128866 | 0.39488459 |
| Un-EV | | | 1.053 | 12.756 | **0.312** | 0.13300950 | 0.12627344 | -0.140320 | 0.40633866 |

**"EV" means the equal variance, abbreviations are used because of insufficient space of table cell.*

To sum up, we corrected Table 1 as follows:

**Table 5: corrected about Table 1**

| | A | B | C | D | E | F | G |
|---|---|---|---|---|---|---|---|
| 1 | Num | $x_i$ | $r_i$ | $y_i$ | $BF_0$ | $BF_{10}$ | $BF_{10}$修 |
| 2 | 1 | 1 | —— | —— | —— | —— | —— |
| 3 | 2 | 1 | ① | ② | —— | —— | —— |
| | | | | …… | | | |
| 91 | 90 | …… | …… | …… | —— | —— | ⑤ |
| | | | | …… | | | |
| 201 | 200 | 0 | …… | …… | ③ | ④ | ⑥ |

Among them, ① and ② have the same meaning as in Table 1.

③：“=(SUM(D$3:D201)+1)/(A201+2)”, it is the Excel expression of formula ⑥. Be careful not to leave out the symbol "$".

④：“=(1-E201)/(E201)”，it is the Excel expression of formula ③

⑤：“=(-0.01416)*F201+1.863972*AVERAGE(B2:B201)^2+(-1.33275)*AVERAGE(B2:B201)+0.35444”, it is the Excel expression of formula ⑧ in the interval of 0.15-0.45.

⑥：“=0.02432*F201+(-2.91139)*AVERAGE(B2:B201)^2+2.205288*AVERAGE(B2:B201)+(-0.45078)”, it is the Excel expression of formula ⑧ in the interval of 0.45-0.55.

**7. Discussion**

To sum up, the overall idea of this paper is to construct another event B based on a Bayesian point estimation algorithm, and prove that B → $H_0$ when n → ∞, so as to convert the solved $P(H_0|x_1,\dots,x_n)$ into a point estimation problem. Finally, through formula ⑧, correct the result to JASP.

However, this paper also leaves a problem: if $\sum_{i=1}^{200} x_i > 110$, what should I do? Because at this time, the Excel values have all been equal to 201, that is to say, the meaning of correction is lost. To this end, we first need to know that when $\sum_{i=1}^{200} x_i$ accounts for n of the total number of samples, Excel will fail? After testing, this value should be about 0.57n. Therefore, we calculated the JASP value at 0.57n, as shown in Table 6.

**Table 6 the BFs of JASP when n>200**

| **n** | **0.57n** | **BFs of JASP** |
|---|---|---|
| 200 | 114 | 0.398 |
| 300 | 171 | 2.381 |
| 325 | 185 | 2.736 |
| 350 | 200 | 4.163 |
| 400 | 228 | 5.545 |
| 450 | 257 | 9.843 |
| 500 | 285 | 13.281 |
| 1000 | 570 | 1282.612 |
| 2000 | 1140 | $1.689{+}10^7$ |

It is obvious that the value of JASP increases significantly with the increase of n. And when $n \in (325, 350)$, the JASP value is approximately greater than 3. Considering that more than value 3 is generally considered to be reliable evidence, so when n>325, such as $r_i>0.57n$, no further verification by JASP is required, which can be considered as strong evidence. When $n \leq 325$, to ensure accuracy, it is recommended to use JASP or R software for verification.

Finally, many preconditions are set for the solution of this paper, such as the estimation premise of unbiased interval is n=200, or the default option of JASP is ">Test value". If these presuppositions are modified, the algorithm idea should be the same. The solution and model modification we left to interested readers.